\documentclass[12pt]{article}

\parskip=\medskipamount            
\def\7#1#2{\mathop{\null#2}\limits^{#1}}        

\def\beee{\begin{equation}}
\def\eeee{\end{equation}}

\begin{document}
\bibliographystyle{unsrt}
 
\begin{center}
\textbf{The Color Charge Degree of Freedom in Particle Physics}\\
[5mm]
O.W. Greenberg\footnote{email address, owgreen@umd.edu.}\\
{Center for Fundamental Physics\\
Department of Physics\\
University of Maryland\\
College Park, MD~~20742-4111, USA\\
University of Maryland Preprint PP-08-04}\\
\end{center}

\begin{abstract}

We review the color charge degree of freedom in particle physics.

\end{abstract}

Color has two facets in particle physics. One is as a three-valued charge degree of
freedom, analogous to electric charge as a degree of freedom in electromagnetism.
The other is as a gauge symmetry, analogous to the $U(1)$ gauge theory of electromagnetism.
Color as a three-valued charge degree of freedom was introduced by 
Oscar W. Greenberg~\cite{owg64} in 1964. Color as a
gauge symmetry was introduced by Yoichiro Nambu~\cite{nam} and by Moo Young Han and 
Yoichiro Nambu~\cite{hannam} in 1965. 
The union of the two contains the essential ingredients of quantum chromodynamics, QCD.
The word ``color'' in this context is purely colloquial and has no connection with 
the color that we see with our eyes in everyday life. 

The theoretical and experimental background to the discovery of color 
centers around events in 1964. In 1964 Murray Gell-Mann~\cite{mgm} and 
George Zweig~\cite{zwe} independently proposed what are
now called ``quarks,'' particles that are constituents of the observed
strongly interacting particles, ``hadrons,'' such as protons and neutrons.
Quarks gave a simple way to account for the quantum numbers of the hadrons.
However quarks were paradoxical in that they had fractional values of their
electric charges, but no such fractionally charged particles had been observed.
Three ``flavors'' of quarks, up, down, and strange, were known at that time. The group
$SU(3)_{flavor}$, acting on these three flavors, gave an approximate symmetry 
that led to mass formulas for the
hadrons constructed with these quarks. However the spin 1/2 of the quarks was not
included in the model. 

The quark spin 1/2 and the symmetry $SU(2)_{spin}$ acting on the two states of spin 1/2
were introduced in the model by Feza G\"ursey and Luigi Radicati~\cite{gur}. They
combined $SU(2)_{spin}$ with $SU(3)_{flavor}$ into a larger
$SU(6)_{spin-flavor}$ symmetry. This larger symmetry unified the previously known mass formulas
for the octet of spin-1/2 baryons and the decuplet of spin-3/2 baryons. Using this
$SU(6)$ theory Mirza A.B. B\'eg, Benjamin W. Lee and Abraham Pais~\cite{beg} calculated 
the ratio of the magnetic moments of the 
proton and neutron to be -3/2, which agrees with experiment to within 3\%. 
However the successful $SU(6)$ theory required that the
configuration of the quarks that gave the correct lowlying baryons must be
in a symmetric state under permutations. This contradicts the spin-statistics theorem
of Wolfgang Pauli~\cite{pau}, according to which quarks as spin-1/2
particles have Fermi statistics and must be in an antisymmetric state under 
permutations.

In the same year 1964 Oscar W. Greenberg~\cite{owg64} recognized that 
this contradiction could be 
resolved by allowing quarks to have a new hidden three-valued charge, 
expressed in terms
of parafermi statistics of order three. This was the discovery of color. 
The antisymmetrization of the hidden
degree of freedom allows the quarks in baryons to be in the observed symmetric configuration
of the visible degrees of freedom: space, spin and flavor. Greenberg called this
model the ``symmetric quark model'' for baryons. As an
observable test of this model, Greenberg constructed a table of the spin,
parity, isospin and strangeness of the orbital excitations of the ground-state
quark configurations in this model. 

In 1964 the hidden color charge on top of the fractionally charged
quarks seemed unduly speculative to some. Independent evidence for 
the existence of color came when
measurements of the properties of excited
baryons confirmed the predictions of the symmetric quark model. It was only in 1968 that
Haim Harari~\cite{har}, as rapporteur for baryon spectrocopy, adopted the symmetric
quark model as the correct model of baryons. 

Additional evidence for color came
from the ratio of the annihilation cross section for $e^+e^-\rightarrow hadrons$
to that for $e^+e^-\rightarrow \mu^+ \mu^-$ and from the decay rate for
$\pi^0 \rightarrow \gamma \gamma$. Both of these follow
from the gauge theory and the parastatistics version of color. 
Further consequences of color
require the gauged theory of color, quantum chromodynamics, QCD, described
below.

In 1965 Yoichiro Nambu~\cite{nam} and, in a separate paper, 
Moo Young Han and Yoichiro Nambu~\cite{hannam} proposed a model with three sets of
quark triplets. Their model has two different $SU(3)$ symmetries. One called $SU(3)^{\prime}$
has the 
original $SU(3)_{flavor}$ symmetry of the quark model and the other, 
called $SU(3)^{\prime \prime}$, makes explicit
the hidden three-valued color charge degree of freedom that had been 
introduced in the
parastatistics model of Greenberg. This model allows the 
$SU(3)^{\prime \prime}$, which can be identified with the present $SU(3)_{color}$ if the
quark charges are chosen fractional, to be
gauged. Indeed Nambu~\cite{nam} and Han and Nambu~\cite{hannam} introduced an
octet of what we now call ``gluons'' as the mediator of the force between the
quarks. The gauging of the three-valued color charge carried by quarks with fractional
electric charges is the present QCD,
the accepted theory of the strong interactions. 

The model of Han and Nambu
assigned integer charges to their three triplets to avoid the fractional electric charges 
of the original quark model. This aspect of the Han-Nambu model 
conflicts both with
experiment and with exact color symmetry and is not part of QCD.
Greenberg and Daniel Zwanziger~\cite{grezwa} made the identity of the 3 of
parafermi statistics of order 3 and the 3 of $SU(3)_{color}$ with
fractionally-charged quarks explicit in 1966.

In addition to the consequences of the parastatistics
model, QCD leads to other important results. These include 
(a) permanent confinement of 
quarks and color, (b) asymptotic freedom, 
discovered by David J. Gross~\cite{gro}, H. David Politzer~\cite{pol} 
and Frank Wilczek~\cite{gro} in 1973, which reconciles 
the low energy behavior of quarks confined in hadrons with the quasi-free behavior of
quarks that interact at high energy and momentum transfer in the parton model, 
(c) running of coupling constants and high-precision tests of QCD at high energy, 
and (d) jets in high energy collisions.

Note: References [1] through [12] are primary references. References [13] through [18]
are secondary references.

\bibliographystyle{unsrt}

\end{document}